\documentclass[aps,12pt,preprint]{revtex4}
\usepackage{graphicx}
\usepackage{amsmath}
\usepackage{hyperref}
\usepackage{bm}
\usepackage[dvips]{color}

\newcommand{\etal} {\textit{et al.}}
\newcommand{\ie} {\textit{i.e.}}
\newcommand{\fzd} {Institute of Ion Beam Physics and Materials Research, Forschungszentrum Dresden-Rossendorf, 01314 Dresden, Germany}

\begin{document}

\title{MnSi$_{1.7}$ nanoparticles embedded in Si: Superparamagnetism with a collective behavior}

\author{Shengqiang~Zhou}
\email[Electronic address: ]{S.Zhou@fzd.de} \affiliation{\fzd}
\author{Artem~Shalimov}
\author{Kay~Potzger}
\author{Manfred~Helm}
\author{J\"{u}rgen~Fassbender}
\author{Heidemarie~Schmidt}
\affiliation{\fzd}

\begin{abstract}
The doping of Mn in Si is attracting research attentions due to the possibility to fabricate Si-based diluted magnetic semiconductors. However,
the low solubility of Mn in Si favors the precipitation of Mn ions even at non-equilibrium growth conditions. MnSi$_{1.7}$ nanoparticles are the
common precipitates, which show exotic magnetic properties in comparison with the MnSi$_{1.7}$ bulk phase. In this paper we present the static
and dynamic magnetic properties of MnSi$_{1.7}$ nanoparticles. Using the Preisach model, we derive the magnetic parameters, such as the
magnetization of individual particles, the distribution of coercive fields and the inter-particle interaction field. Time-dependent
magnetization measurements reveal aging and memory effects, qualitatively similar to those seen in spin glasses .
\end{abstract}
\maketitle




\section{Introduction}\label{introduction}

Diluted magnetic semiconductors (DMS) are attracting great interest because of their potential use for spintronics. GaAs:Mn has recently emerged
as the most popular material for this new technology. However, Si-based DMS would be preferably used because of the availability of high
quality, large-size, and low-cost wafers. More importantly, the fabrication of Si-based DMS is compatible with the mature microelectronics
technique. Based on the Zener model, Dietl \emph{et al.} \cite{dietl00} predicted carrier-mediated ferromagnetism if p-type conducting Si is
doped with 5\% Mn. Using density-functional theory, Wu \emph{et al.} \cite{wu:117202} demonstrated that interstitial Mn can be utilized to tune
the magnetic properties of Si. Experimentally, various groups have reported the observation of ferromagnetism in Mn doped Si
\cite{zhang:786,bolduc:033302,zhou07si,ko:033912,yabuuchi:045307,JJAP.47.4487}. The reported Curie temperatures range from 200 to 400 K.
However, the opinions concerning the origin of the observed ferromagnetism are very diverse. Using high resolution, spatially resolved
techniques, comprehensive material characterization reveals the clustering of Mn-rich phases in Mn implanted Si, namely MnSi$_{1.7}$
\cite{zhou07si,ko:033912,ko:053912,awo-affouda:1644,JJAP.47.4487}, which is the energetically most favorable Mn-silicide phase
\cite{zou:133111,Mn-silicide_nanowire}. Moverover, Mn-rich phases also form during pulsed laser annealing following Mn ion implantation into Si
\cite{Peng2009}. However, Mn-precipitations in Mn implanted GaAs are successfully suppressed \cite{scarpulla03}. Nevertheless, after considering
the formation of MnSi$_{1.7}$ nanoparticles exotic magnetic properties have been observed. The magnetization per Mn is as large as 0.21
$\mu_B$/Mn, which is much larger than that (0.012 $\mu_B$/Mn) of bulk MnSi$_{1.7}$ \cite{zhou07si}. Ko \emph{et al.} suggested the existence of
multi-fold contributions to the observed ferromagnetism according to their observation of Mn-rich and Mn-poor phases \cite{ko:033912,ko:053912}.
It is important to understand the dynamic magnetization of an ensemble of ferromagnetic nanoparticles due to its influence on technological
applications. Spin-glass-like behavior has been observed in GaAs:Mn and Ge:Mn systems containing Mn-rich clusters
\cite{wang:202503,jaeger:045330,zhou_Mn5Ge3}. Despite numerous publications on the structural and magnetic properties of MnSi$_{1.7}$ embedded
in Si \cite{zhou07si,ko:033912,yabuuchi:045307,JJAP.47.4487,BakMisiuk200999}, information of their dynamic properties is lacking. In this paper,
we will present the static and dynamic magnetic properties of MnSi$_{1.7}$ nanoparticles. The magnetic parameters, such as the magnetization of
individual particles, the distribution of coercive fields and the inter-particle interaction, are deduced using the Preisach model
\cite{Preisach,shalimov:064906}. Time-dependent magnetization measurements reveal a spin-glass-like behavior.

\section{Experimental methods}

The Mn-implanted Si samples were prepared from commercially available, Czochralski grown single-crystal Si(001) wafers, which were p-type doped
with a B concentration of $1.2\times10^{19}$ cm$^{-3}$. Mn$^+$ ions were implanted at an energy of 300 keV with a fluence of $1.0\times10^{16}$
cm$^{-2}$, which corresponds to a peak concentration of 0.8\%, with a projected range (R$_p$) of 258$\pm$82 nm. The samples were held at 350
$^{\circ}$C during implantation to avoid amorphization. In order to reduce channeling effects, the angle between the sample surface normal and
the incident beam was set to 7$^{\circ}$. After implantation, rapid thermal annealing (RTA) was performed at a temperature of 800 $^{\circ}$C
for 5 min in a forming gas of N$_2$. Magnetic properties were analyzed using a superconducting quantum interference device (SQUID) magnetometer
(Quantum Design MPMS). In order to obtain the static and dynamic magnetic properties, we measured the magnetization depending on field,
temperature and time.

\section{Results and discussion}\label{results}
\subsection{Static magnetic properties}

\begin{figure} \center
\includegraphics[scale=1.2]{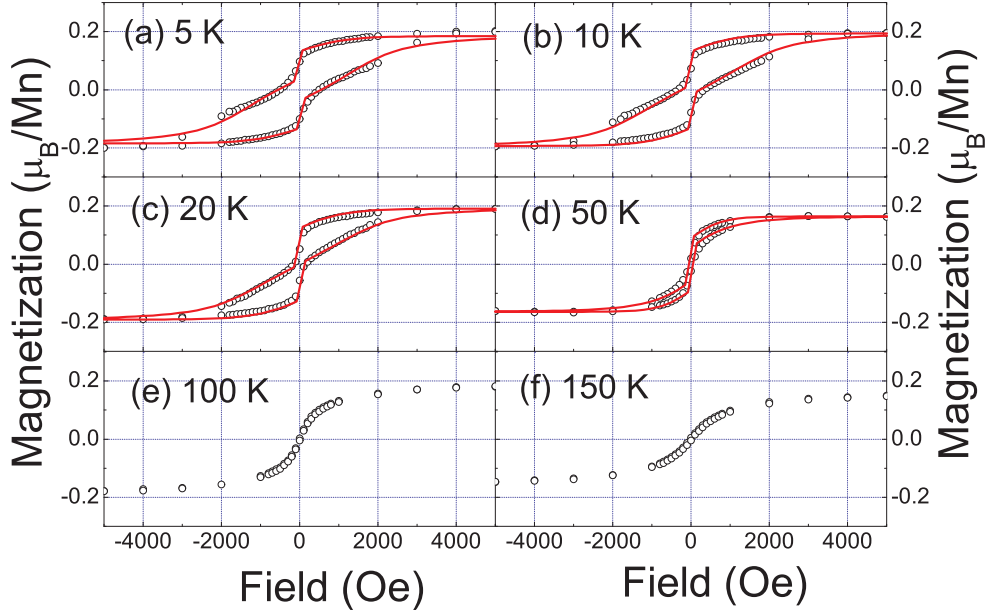}
\caption{Hysteresis loops (open circles) measured in the temperature range from 5 K to 150 K. The solid curves
(a-d) are fittings using the Preisach model.}\label{fig:MH}
\end{figure}

\begin{figure} \center
\includegraphics[scale=0.8]{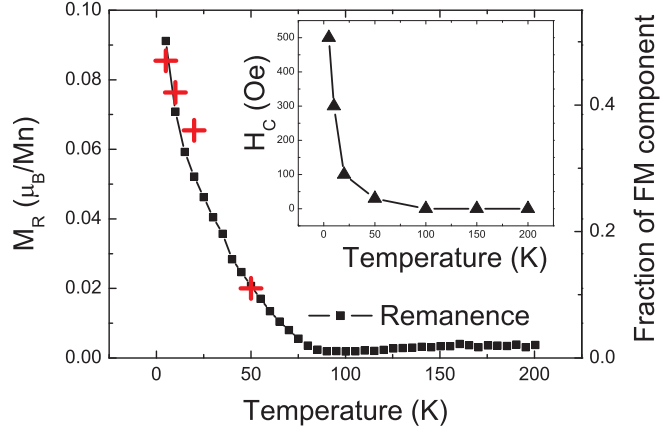}
\caption{Measured temperature dependent remanence and coercivity (inset). Both drop to zero at the temperature
around 80 K. The red crosses display the fraction of the ferromagnetic components obtained by fitting using the
Preisach model. Both the remanence and the fraction of ferromagnetic component have the same temperature
dependence. }\label{fig:rem_Hc}
\end{figure}

The structural properties of the sample after rapid thermal annealing have been reported in Ref. \onlinecite{zhou07si}. MnSi$_{1.7}$
nanoparticles are formed with the average diameter of 11 nm with a sphere like shape. We measured the magnetization of the sample by applying
the field parallel or perpendicular to the sample surface and did not find any difference, which hints towards an isotropic magnetization.
However, the same effect would be seen from a collection of anisotropic particles with randomly-oriented easy axis. The M-H data for the sample
with the field applied in the film plane is shown in Figure \ref{fig:MH}. The linear diamagnetic background of the Si wafer has been subtracted
for all shown data. The magnetization is normalized by the number of implanted Mn ions. It can be seen that the sample exhibits a hysteretic
behavior with the magnetic remanence being 33\% of the saturation magnetization at 5 K. With increasing temperature, both the coercivity
(\emph{H}$_C$) and remanence (\emph{M}$_R$) drop rapidly. At 100 K and 150 K, the coercivity and remanence are zero. This is the evidence for
superparamagnetism. The blocking temperature lies between 50 K and 100 K. Figure \ref{fig:rem_Hc} shows the remanent magnetization measured from
5 to 200 K after shutting down the field of 10000 Oe at 5 K. The remanence decreases rapidly with increasing temperature and falls to zero at
around 80 K. The coercivity (inset of Figure \ref{fig:rem_Hc}) exhibits a similar temperature dependence. Note that bulk MnSi$_{1.7}$ is
reported to exhibit weak itinerant magnetism with an ordering temperature of 47 K and with a very low saturation moment of 0.012 $\mu_B$/Mn
\cite{gottlieb03}, being much different from the MnSi$_{1.7}$ nanoparticles investigated here. In our samples, the largest magnetization is 0.21
$\mu_B$/Mn. Independently, Yabuuchi \emph{et al.} \cite{JJAP.47.4487} used slightly different implantation and annealing conditions and realized
a maximum magnetization of 0.2 $\mu_B$/Mn. Both values are much larger than for a bulk MnSi$_{1.7}$ crystal. Using first-principles calculation,
Yabuuchi \emph{et al.} have clarified that the stoichiometry, strain and charge accumulation as well as the interface between MnSi$_{1.7}$ and
Si strongly influence the magnetic properties of MnSi$_{1.7}$ nanoparticles \cite{yabuuchi:045307}. These effects well account for the
experimental observations.

\begin{figure} \center
\includegraphics[scale=0.8]{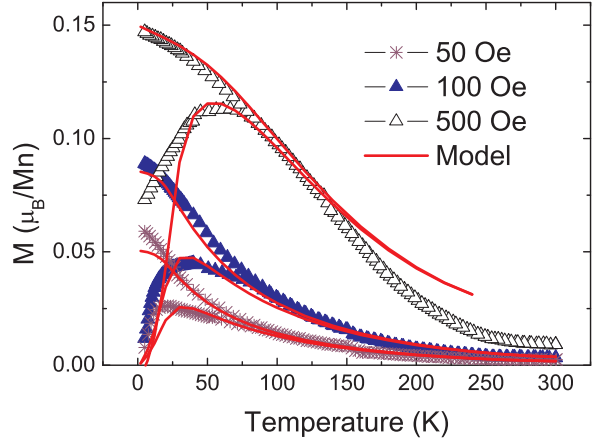}
\caption{Field-cooled (FC, upper branches) and zero-field-cooled (ZFC, lower branches) M-T curves measured at different fields. The solid lines
are fitting lines using the Preisach model.}\label{fig:ZFCFC}
\end{figure}

The temperature dependent magnetization under zero field cooling and field cooling was measured in the following way. The sample was cooled in
zero field from above room temperature to 5 K. Then a field was applied and the zero-field cooled (ZFC) magnetization curve was measured with
increasing temperature from 5 to 300 K, after which the field-cooled (FC) magnetization curve was measured in the same field from 300 to 5 K
with decreasing temperature. Figure \ref{fig:ZFCFC} shows the ZFC/FC magnetization curves measured at different fields. An irreversible behavior
is observed in ZFC/FC curves. Such an irreversibility originates from the anisotropy barrier blocking of the magnetization orientation in the
nanoparticles cooled under zero field. The magnetization direction of the nanoparticles is frozen in its initial state at high temperature, \ie,
randomly oriented. At low temperature (5 K in our case), a magnetic field is applied. Some small nanoparticles with small magnetic anisotropy
energy flip along the field direction, while the larger ones do not. With increasing temperature, the thermal activation energy together with
the field also flips the larger particles. This process results in the increase of the magnetization in the ZFC curve with temperature. The peak
of the ZFC curves, \emph{T}$_P$, is considered as the average blocking temperature of the sample. A notable feature is the increase of
\emph{T}$_P$ with increasing applied field. Such a field dependence of T$_P$ has been observed in several magnetic nanoparticle systems, such as
Fe$_3$O$_4$ \cite{PhysRevLett.67.2721}, $\gamma$-Fe$_2$O$_3$ \cite{PhysRevB.56.14551}, ferritin \cite{PhysRevB.56.10793} and FePt
\cite{zhengrk}. The physical origin of this behavior is discussed concerning the size distribution of nanoparticles \cite{zhengrk} and the
inter-particle interaction \cite{PhysRevB.56.10793}.

\begin{figure} \center
\includegraphics[scale=0.8]{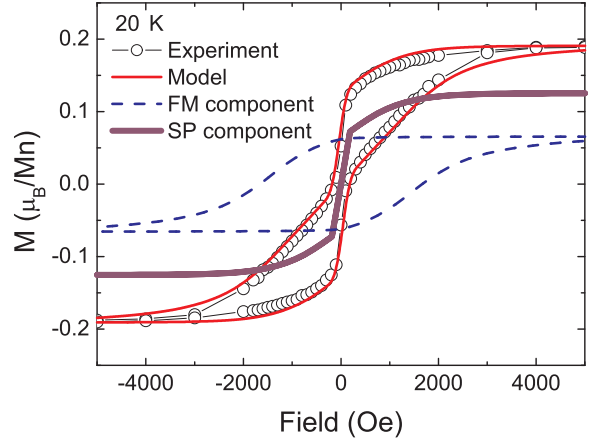}
\caption{Fitting of the hysteresis loop measured at 20 K by the Preisach model. The superparamagnetic (SP) and ferromagnetic (FM) components are
shown in the figure separately.}\label{fig:loop_20K}
\end{figure}

We analyzed the experimental curves using simulations based on the Preisach approach \cite{Song_preisach,shalimov:064906} and derived the key
magnetic parameters of MnSi$_{1.7}$ nanoparticles, namely the magnetization of individual particles and the distributions of coercive and
inter-particle interaction fields. The temperature dependence of the parameters $p$ describing the magnetic properties of the nanoparticles is
usually expressed by the critical temperature \emph{T}$_C$ and the critical exponent $\gamma$
\begin{equation}\label{praisach}
p=p_{0}(1-T/T_{C})^{\gamma},
\end{equation} with $p$ (The subscript 0 denotes the parameters at 0 K) substituted by the mean
magnetic moment $\mu$ of an individual cluster (where $\gamma=\Gamma$), the mean coercivity \emph{H}$_C$ ($\gamma=\Gamma_c$), or the dispersions
$\sigma_i$ ($\gamma=\Gamma_i$) and $\sigma_c$ ($\gamma=\Gamma_c'$) of the interparticle interaction and coercive field, respectively. Long-range
interaction field was equal to zero. The fitting parameters are shown in Tab. \ref{tab:sample}.

Figure \ref{fig:loop_20K} shows a representative fitting of the hysteresis measured at 20 K. Two components contribute to the measured
magnetization: a superparamagnetic (SP) component and a ferromagnetic (FM) component with a large coercive field of 1100 Oe. Note that the
modelled coercivity of the FM component is larger than that shown in the inset of Figure \ref{fig:rem_Hc}, and the modelled and the measured
remanence are the same. The smaller coercivity seen in the measured hysteresis is due to the superposition of the FM and SP components. The
fraction of the ferromagnetic component rapidly decreases with increasing temperature, and is plotted in Figure \ref{fig:rem_Hc}. It is worth to
note that using the same set of fitting parameters we are able to fit the experimental curves measured at different temperatures (see Figure
\ref{fig:MH}).

The average magnetic moment of individual particles $\mu_0$ (at 0 K) computed from the fitting is 4.0$\times$10$^{-17}$ emu (around 4300
$\mu_B$). The average diameter of MnSi$_{1.7}$ particles in this sample is around 11 nm \cite{zhou07si}. In one particle there are approx. 21000
Mn atoms, and we obtain 0.204 $\mu_B$/Mn. This value is in a good agreement with the experiment (0.21 $\mu_B$/Mn). In the work by Yabuuchi
\emph{et al.} \cite{JJAP.47.4487}, the samples with the same Mn fluence were annealed at different temperatures to tune the size of MnSi$_{1.7}$
particles. The ferromagnetism strongly depends on the average particle diameters and is maximized after annealing at 750 $^{\circ}$C. This
annealing temperature resulted in an average diameter of 10 nm for MnSi$_{1.7}$ particles. By the two experimental works
\cite{zhou07si,JJAP.47.4487} and our fitting using the Preisach approach, we confirm that the ferromagnetism in MnSi$_{1.7}$ particles strongly
depends on their diameters: \emph{i.e.}, the exotic ferromagnetism in MnSi$_{1.7}$ particles is mainly a size effect \cite{yabuuchi:045307}.

\begin{table*}
\caption{\label{tab:sample} Best fit Preisach parameters.}
\begin{ruledtabular}
\begin{tabular}{cccccccccc}
T$_C$ (K) & $\mu_0$ (emu) & N (cm$^{-2}$) & $\Gamma$ & $\Gamma_c$ &\emph{H}$_{c0}$ (Oe) & $\sigma _{c0}$ (Oe) & $\Gamma_c'$ & $\sigma _{i0}$ (Oe) & $\Gamma_i$ \\
\hline
320 & 4.0$\times$10$^{-17}$ & 4.3 $\times$10$^{11}$ & 0.6 & 0.6 & 1500 & 2000 & 0.6 & 200 & 0.1\\
\end{tabular}
\end{ruledtabular}
\end{table*}

The ferromagnetic component has a large coercive field of 1500 Oe at 0 K, while the inter-particle interaction is weak. Note that $\sigma_i$ is
around 200 Oe and much smaller than the coercive field (Tab. \ref{tab:sample}). These features are important for practical applications. Using
the same fitting parameters, the ZFC/FC curves are also well fitted as shown in Figure \ref{fig:ZFCFC}. Importantly, the fitting reproduces the
field dependence of \emph{T}$_P$ in the ZFC curves well. Indeed, Song \emph{et al.} have discussed the influence of the inter-particle
interaction using the Preisach model \cite{Song200024}. In the limit of weak interactions \emph{T}$_P$ increases monotonically with applied
fields, while in the limit of strong interactions \emph{T}$_P$ decreases monotonically with applied fields. Therefore, we attribute the increase
of \emph{T}$_P$ at larger fields to the weak interaction between MnSi$_{1.7}$ particles.

\subsection{Dynamic magnetic properties}

To obtain the dynamic magnetic properties of the system, we performed time-dependent measurements. The time-dependent thermoremanent
magnetization (TRM) was measured below the blocking temperature. TRM is measured by cooling the sample in an applied field from an initial
temperature above any spin glass transition to some final temperature, decreasing the field to zero and observing the decaying remanent
magnetization ($M_{r}(t)$). In our case, TRM data were taken after cooling from 300 K to 5 K and 20 K, respectively, in an applied field of 100
Oe.

\begin{figure} \center
\includegraphics[scale=1.3]{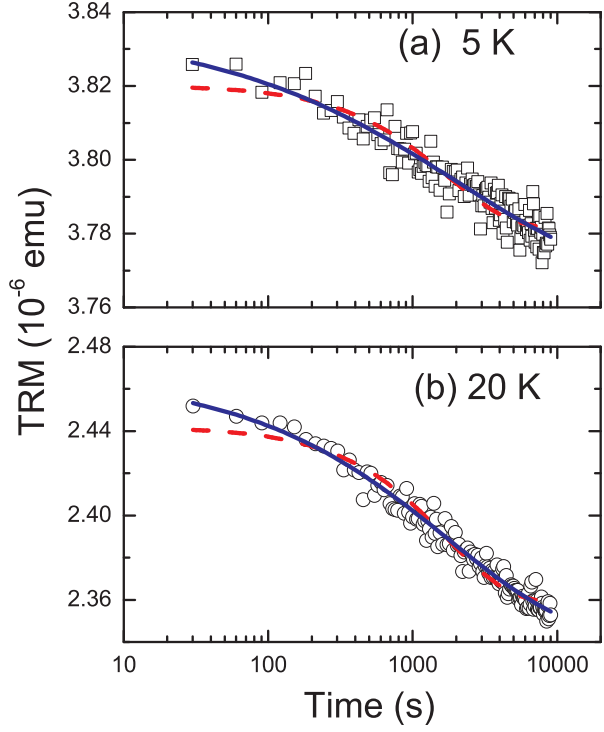}
\caption{TRM time-decays of the field-cooled magnetization (H = 100 Oe) at (a) 5 K and (b) 20 K. Scattered symbols are experimental data. Solid
lines (blue) are stretched-exponential fits with parameters of $\tau$=2120 s and $b$=0.44 at 5 K, and $\tau$=1806 s and $b$=0.53 at 20 K. Dashed
lines (red) are first-order exponential fits with parameters of $\tau$=1721.2 s at 5 K and $\tau$=1720.5 s at 20 K.}\label{fig:relaxation}
\end{figure}

For superparamagnetic nanoparticles, the magnetization decay is usually exponential to first order, such that
one should observe a single relaxation rate approximation \cite{PhysRevB.52.12779}:
\begin{equation}\label{relaxation_exp}
    M_{r}(t)=M_{0}e^{-t/\tau}+C,
\end{equation} where $M_0$ is the initial magnetization, $\tau$ is the relaxation time and $C$ is a constant.
If the superparamagnetic nanoparticles undergo collective behavior due to direct dipole-dipole interaction or a
large particle-size distribution, a stretched exponential form is expected \cite{jaeger:045330}.
\begin{equation}\label{relaxation_stretched}
    M_{r}(t)=M_{0}e^{-(t/\tau)^b}+C,
\end{equation} where $b$ affects the relaxation rate of the glassy component. Figure \ref{fig:relaxation} shows the TRM time-decays at 5 and 20 K.
The stretched exponential relaxation fits better to the experimental data. The fitted parameters of
$\tau$ and $b$ are in the typical range of a spin-glass system \cite{jaeger:045330}.

\begin{figure} \center
\includegraphics[scale=0.8]{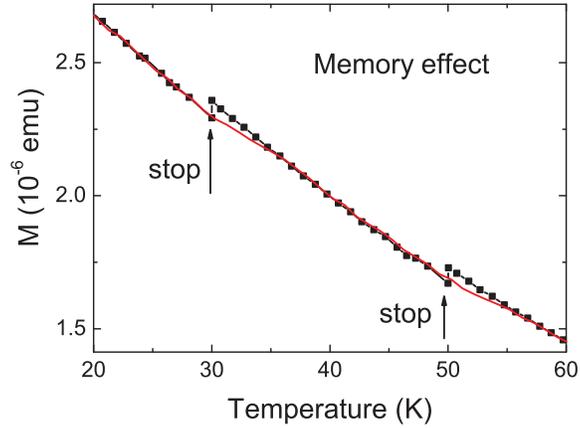}
\caption{Temperature dependent memory effect in the dc magnetization. The solid squares are measured during cooling in a field of 100 Oe at the
same rate but with a stop of 2 hours at 50 K and 30 K, respectively. The field is cut off during stop. The solid line is measured with
continuous heating at the same rate after the previous cooling protocol.}\label{fig:memory}
\end{figure}

In order to further confirm the glass behavior of the system, we also performed history-dependent magnetic memory measurements using the cooling
and heating protocol suggested by Sun \etal~\cite{PhysRevLett.91.167206}. We cooled the sample at 100 Oe and recorded the magnetization during
cooling, but temporarily stopped at \emph{T} = 50 K and 30 K for a period of 2 hours. During the waiting period, the field was switched off.
After the stop, the 100 Oe field was reapplied and cooling and measuring were resumed. The temporary stop resulted in a steplike \emph{M(T)}
curve (solid squares in Figure \ref{fig:memory}). After reaching the lowest temperature 5 K, the sample was heated back in the same field, and
the magnetization was recorded again. The \emph{M(T)} curve during this heating also has a steplike behavior at the stop temperature of 30 and
50 K, then recovers the previous \emph{M(T)} curve measured during cooling. The system remembers its thermal history.

The observed memory effect as well as the relaxation phenomena show qualitative similarities to the spin glass behavior. Two explanations have
been suggested \cite{sasaki:104405}. The first one is a broad distribution of blocking temperatures originating from the distribution of the
anisotropy energy barriers. Another explanation is the strong dipolar interaction between nanoparticles, which frustrates the nanomagnetic
moments, and slows down their relaxation. Our observations support the first model. First of all, using the Preisach model we derived a small
interaction between MnSi$_{1.7}$ nanoparticles. Second, the size of MnSi$_{1.7}$ nanoparticles approximately amounts to 11$\pm$5 nm according to
the TEM observation as shown in Ref. \onlinecite{zhou07si}. The spin flip time for magnetic particles depends exponentially on the particle
size. Therefore, even a small distribution of the particle size could give a broad range of relaxation times. Therefore, the observed
spin-glass-like behavior is attributed to the broad distribution of particle sizes, \ie~of anisotropy energy barriers.

\section{Conclusions}\label{conclusions}

Static and dynamic magnetic properties of Mn implanted Si were investigated. The magnetic properties of Mn-silicide nanoparticles can be well
explained using the Preisach model. We found that there are two components (superparamagnetic and ferromagnetic fractions) contributing to the
magnetism. The fraction of the ferromagnetic component decreases with increasing measurement temperature. The interaction between the magnetic
nanoparticles is weak. Therefore, the superparamagnetism blocking-temperature increases monotonically with applied field. Time-dependent
measurements, \emph{i.e.} relaxation and memory effect, support a spin-glass-like behavior in the investigated material system, which results
from the size distribution of MnSi$_{1.7}$ nanoparticles.

\section{Acknowledgement}

The author (S.Z.) acknowledges the financial funding from the Bundesministerium f\"{u}r Bildung und Forschung (FKZ13N10144), and A.S. wants to
thank the Deutsche Forschungsgemeinschaft (PO1275/2-1, 'SEMAN').


\begin{thebibliography}{30}
\expandafter\ifx\csname natexlab\endcsname\relax\def\natexlab#1{#1}\fi \expandafter\ifx\csname
bibnamefont\endcsname\relax
  \def\bibnamefont#1{#1}\fi
\expandafter\ifx\csname bibfnamefont\endcsname\relax
  \def\bibfnamefont#1{#1}\fi
\expandafter\ifx\csname citenamefont\endcsname\relax
  \def\citenamefont#1{#1}\fi
\expandafter\ifx\csname url\endcsname\relax
  \def\url#1{\texttt{#1}}\fi
\expandafter\ifx\csname urlprefix\endcsname\relax\def\urlprefix{URL }\fi \providecommand{\bibinfo}[2]{#2}
\providecommand{\eprint}[2][]{\url{#2}}

\bibitem[{\citenamefont{Dietl et~al.}(2000)\citenamefont{Dietl, Ohno,
  Matsukura, Cibert, and Ferrand}}]{dietl00}
\bibinfo{author}{\bibfnamefont{T.}~\bibnamefont{Dietl}},
  \bibinfo{author}{\bibfnamefont{H.}~\bibnamefont{Ohno}},
  \bibinfo{author}{\bibfnamefont{F.}~\bibnamefont{Matsukura}},
  \bibinfo{author}{\bibfnamefont{J.}~\bibnamefont{Cibert}}, \bibnamefont{and}
  \bibinfo{author}{\bibfnamefont{D.}~\bibnamefont{Ferrand}},
  \bibinfo{journal}{Science} \textbf{\bibinfo{volume}{287}},
  \bibinfo{pages}{1019} (\bibinfo{year}{2000}).

\bibitem[{\citenamefont{Wu et~al.}(2007)\citenamefont{Wu, Kratzer, and
  Scheffler}}]{wu:117202}
\bibinfo{author}{\bibfnamefont{H.}~\bibnamefont{Wu}},
  \bibinfo{author}{\bibfnamefont{P.}~\bibnamefont{Kratzer}}, \bibnamefont{and}
  \bibinfo{author}{\bibfnamefont{M.}~\bibnamefont{Scheffler}},
  \bibinfo{journal}{Phys. Rev. Lett.} \textbf{\bibinfo{volume}{98}},
  \bibinfo{pages}{117202} (\bibinfo{year}{2007}).

\bibitem[{\citenamefont{Zhang et~al.}(2004)\citenamefont{Zhang, Liu, Gao, Wu,
  Du, Zhu, Xiao, and Chen}}]{zhang:786}
\bibinfo{author}{\bibfnamefont{F.~M.} \bibnamefont{Zhang}},
  \bibinfo{author}{\bibfnamefont{X.~C.} \bibnamefont{Liu}},
  \bibinfo{author}{\bibfnamefont{J.}~\bibnamefont{Gao}},
  \bibinfo{author}{\bibfnamefont{X.~S.} \bibnamefont{Wu}},
  \bibinfo{author}{\bibfnamefont{Y.~W.} \bibnamefont{Du}},
  \bibinfo{author}{\bibfnamefont{H.}~\bibnamefont{Zhu}},
  \bibinfo{author}{\bibfnamefont{J.~Q.} \bibnamefont{Xiao}}, \bibnamefont{and}
  \bibinfo{author}{\bibfnamefont{P.}~\bibnamefont{Chen}},
  \bibinfo{journal}{Appl. Phys. Lett.} \textbf{\bibinfo{volume}{85}},
  \bibinfo{pages}{786} (\bibinfo{year}{2004}).

\bibitem[{\citenamefont{Bolduc et~al.}(2005)\citenamefont{Bolduc, Awo-Affouda,
  Stollenwerk, Huang, Ramos, Agnello, and LaBella}}]{bolduc:033302}
\bibinfo{author}{\bibfnamefont{M.}~\bibnamefont{Bolduc}},
  \bibinfo{author}{\bibfnamefont{C.}~\bibnamefont{Awo-Affouda}},
  \bibinfo{author}{\bibfnamefont{A.}~\bibnamefont{Stollenwerk}},
  \bibinfo{author}{\bibfnamefont{M.~B.} \bibnamefont{Huang}},
  \bibinfo{author}{\bibfnamefont{F.~G.} \bibnamefont{Ramos}},
  \bibinfo{author}{\bibfnamefont{G.}~\bibnamefont{Agnello}}, \bibnamefont{and}
  \bibinfo{author}{\bibfnamefont{V.~P.} \bibnamefont{LaBella}},
  \bibinfo{journal}{Phys. Rev. B} \textbf{\bibinfo{volume}{71}},
  \bibinfo{pages}{033302} (\bibinfo{year}{2005}).

\bibitem[{\citenamefont{Zhou et~al.}(2007)\citenamefont{Zhou, Potzger, Zhang,
  M\"{u}cklich, Eichhorn, Schell, Gr\"{o}tzchel, Schmidt, Skorupa, Helm
  et~al.}}]{zhou07si}
\bibinfo{author}{\bibfnamefont{S.}~\bibnamefont{Zhou}},
  \bibinfo{author}{\bibfnamefont{K.}~\bibnamefont{Potzger}},
  \bibinfo{author}{\bibfnamefont{G.}~\bibnamefont{Zhang}},
  \bibinfo{author}{\bibfnamefont{A.}~\bibnamefont{M\"{u}cklich}},
  \bibinfo{author}{\bibfnamefont{F.}~\bibnamefont{Eichhorn}},
  \bibinfo{author}{\bibfnamefont{N.}~\bibnamefont{Schell}},
  \bibinfo{author}{\bibfnamefont{R.}~\bibnamefont{Gr\"{o}tzchel}},
  \bibinfo{author}{\bibfnamefont{B.}~\bibnamefont{Schmidt}},
  \bibinfo{author}{\bibfnamefont{W.}~\bibnamefont{Skorupa}},
  \bibinfo{author}{\bibfnamefont{M.}~\bibnamefont{Helm}},
  \bibinfo{author}{\bibfnamefont{J.}~\bibnamefont{Fassbender}},
  \bibinfo{author}{\bibfnamefont{D.}~\bibnamefont{Geiger}},
  \bibinfo{journal}{Phys. Rev. B} \textbf{\bibinfo{volume}{75}},
  \bibinfo{pages}{085203} (\bibinfo{year}{2007}).

\bibitem[{\citenamefont{Ko et~al.}(2008{\natexlab{a}})\citenamefont{Ko, Teo,
  Liew, Chong, MacKenzie, MacLaren, and Chapman}}]{ko:033912}
\bibinfo{author}{\bibfnamefont{V.}~\bibnamefont{Ko}},
  \bibinfo{author}{\bibfnamefont{K.~L.} \bibnamefont{Teo}},
  \bibinfo{author}{\bibfnamefont{T.}~\bibnamefont{Liew}},
  \bibinfo{author}{\bibfnamefont{T.~C.} \bibnamefont{Chong}},
  \bibinfo{author}{\bibfnamefont{M.}~\bibnamefont{MacKenzie}},
  \bibinfo{author}{\bibfnamefont{I.}~\bibnamefont{MacLaren}}, \bibnamefont{and}
  \bibinfo{author}{\bibfnamefont{J.~N.} \bibnamefont{Chapman}},
  \bibinfo{journal}{J. Appl. Phys.} \textbf{\bibinfo{volume}{104}},
  \bibinfo{pages}{033912} (\bibinfo{year}{2008}{\natexlab{a}}).

\bibitem[{\citenamefont{Yabuuchi
  et~al.}(2008{\natexlab{a}})\citenamefont{Yabuuchi, Kageshima, Ono, Nagase,
  Fujiwara, and Ohta}}]{yabuuchi:045307}
\bibinfo{author}{\bibfnamefont{S.}~\bibnamefont{Yabuuchi}},
  \bibinfo{author}{\bibfnamefont{H.}~\bibnamefont{Kageshima}},
  \bibinfo{author}{\bibfnamefont{Y.}~\bibnamefont{Ono}},
  \bibinfo{author}{\bibfnamefont{M.}~\bibnamefont{Nagase}},
  \bibinfo{author}{\bibfnamefont{A.}~\bibnamefont{Fujiwara}}, \bibnamefont{and}
  \bibinfo{author}{\bibfnamefont{E.}~\bibnamefont{Ohta}},
  \bibinfo{journal}{Phys. Rev. B} \textbf{\bibinfo{volume}{78}},
  \bibinfo{pages}{045307} (\bibinfo{year}{2008}{\natexlab{a}}).

\bibitem[{\citenamefont{Yabuuchi
  et~al.}(2008{\natexlab{b}})\citenamefont{Yabuuchi, Ono, Nagase, Kageshima,
  Fujiwara, and Ohta}}]{JJAP.47.4487}
\bibinfo{author}{\bibfnamefont{S.}~\bibnamefont{Yabuuchi}},
  \bibinfo{author}{\bibfnamefont{Y.}~\bibnamefont{Ono}},
  \bibinfo{author}{\bibfnamefont{M.}~\bibnamefont{Nagase}},
  \bibinfo{author}{\bibfnamefont{H.}~\bibnamefont{Kageshima}},
  \bibinfo{author}{\bibfnamefont{A.}~\bibnamefont{Fujiwara}}, \bibnamefont{and}
  \bibinfo{author}{\bibfnamefont{E.}~\bibnamefont{Ohta}}, \bibinfo{journal}{J.
  J. Appl. Phys.} \textbf{\bibinfo{volume}{47}}, \bibinfo{pages}{4487}
  (\bibinfo{year}{2008}{\natexlab{b}}).

\bibitem[{\citenamefont{Ko et~al.}(2008{\natexlab{b}})\citenamefont{Ko, Teo,
  Liew, Chong, Liu, Wee, Du, Stoffel, and Schmidt}}]{ko:053912}
\bibinfo{author}{\bibfnamefont{V.}~\bibnamefont{Ko}},
  \bibinfo{author}{\bibfnamefont{K.~L.} \bibnamefont{Teo}},
  \bibinfo{author}{\bibfnamefont{T.}~\bibnamefont{Liew}},
  \bibinfo{author}{\bibfnamefont{T.~C.} \bibnamefont{Chong}},
  \bibinfo{author}{\bibfnamefont{T.}~\bibnamefont{Liu}},
  \bibinfo{author}{\bibfnamefont{A.~T.~S.} \bibnamefont{Wee}},
  \bibinfo{author}{\bibfnamefont{A.~Y.} \bibnamefont{Du}},
  \bibinfo{author}{\bibfnamefont{M.}~\bibnamefont{Stoffel}}, \bibnamefont{and}
  \bibinfo{author}{\bibfnamefont{O.~G.} \bibnamefont{Schmidt}},
  \bibinfo{journal}{J. Appl. Phys.} \textbf{\bibinfo{volume}{103}},
  \bibinfo{eid}{053912} (\bibinfo{year}{2008}{\natexlab{b}}).

\bibitem[{\citenamefont{Awo-Affouda et~al.}(2006)\citenamefont{Awo-Affouda,
  Bolduc, Huang, Ramos, Dunn, Thiel, Agnello, and LaBella}}]{awo-affouda:1644}
\bibinfo{author}{\bibfnamefont{C.}~\bibnamefont{Awo-Affouda}},
  \bibinfo{author}{\bibfnamefont{M.}~\bibnamefont{Bolduc}},
  \bibinfo{author}{\bibfnamefont{M.~B.} \bibnamefont{Huang}},
  \bibinfo{author}{\bibfnamefont{F.}~\bibnamefont{Ramos}},
  \bibinfo{author}{\bibfnamefont{K.~A.} \bibnamefont{Dunn}},
  \bibinfo{author}{\bibfnamefont{B.}~\bibnamefont{Thiel}},
  \bibinfo{author}{\bibfnamefont{G.}~\bibnamefont{Agnello}}, \bibnamefont{and}
  \bibinfo{author}{\bibfnamefont{V.~P.} \bibnamefont{LaBella}},
  \bibinfo{journal}{J. Vac. Sci. Technol. A} \textbf{\bibinfo{volume}{24}},
  \bibinfo{pages}{1644} (\bibinfo{year}{2006}).

\bibitem[{\citenamefont{Zou et~al.}(2007)\citenamefont{Zou, Wang, Wang, Wang,
  Mao, and Kong}}]{zou:133111}
\bibinfo{author}{\bibfnamefont{Z.-Q.} \bibnamefont{Zou}},
  \bibinfo{author}{\bibfnamefont{H.}~\bibnamefont{Wang}},
  \bibinfo{author}{\bibfnamefont{D.}~\bibnamefont{Wang}},
  \bibinfo{author}{\bibfnamefont{Q.-K.} \bibnamefont{Wang}},
  \bibinfo{author}{\bibfnamefont{J.-J.} \bibnamefont{Mao}}, \bibnamefont{and}
  \bibinfo{author}{\bibfnamefont{X.-Y.} \bibnamefont{Kong}},
  \bibinfo{journal}{Appl. Phys. Lett.} \textbf{\bibinfo{volume}{90}},
  \bibinfo{pages}{133111} (\bibinfo{year}{2007}).

\bibitem[{\citenamefont{Wang and Zou}(2009)}]{Mn-silicide_nanowire}
\bibinfo{author}{\bibfnamefont{D.}~\bibnamefont{Wang}} \bibnamefont{and}
  \bibinfo{author}{\bibfnamefont{Z.-Q.} \bibnamefont{Zou}},
  \bibinfo{journal}{Nanotechnology} \textbf{\bibinfo{volume}{20}},
  \bibinfo{pages}{275607} (\bibinfo{year}{2009}).

\bibitem[{\citenamefont{Peng et~al.}(2009)\citenamefont{Peng, Jeynes, Bailey,
  Adikaari, Stolojan, and Webb}}]{Peng2009}
\bibinfo{author}{\bibfnamefont{N.}~\bibnamefont{Peng}},
  \bibinfo{author}{\bibfnamefont{C.}~\bibnamefont{Jeynes}},
  \bibinfo{author}{\bibfnamefont{M.~J.} \bibnamefont{Bailey}},
  \bibinfo{author}{\bibfnamefont{D.}~\bibnamefont{Adikaari}},
  \bibinfo{author}{\bibfnamefont{V.}~\bibnamefont{Stolojan}}, \bibnamefont{and}
  \bibinfo{author}{\bibfnamefont{R.~P.} \bibnamefont{Webb}},
  \bibinfo{journal}{Nucl. Instr. and Meth. B} \textbf{\bibinfo{volume}{267}},
  \bibinfo{pages}{1623} (\bibinfo{year}{2009}).

\bibitem[{\citenamefont{Scarpulla et~al.}(2003)\citenamefont{Scarpulla, Dubon,
  Yu, Monteiro, Pillai, Aziz, and Ridgway}}]{scarpulla03}
\bibinfo{author}{\bibfnamefont{M.~A.} \bibnamefont{Scarpulla}},
  \bibinfo{author}{\bibfnamefont{O.~D.} \bibnamefont{Dubon}},
  \bibinfo{author}{\bibfnamefont{K.~M.} \bibnamefont{Yu}},
  \bibinfo{author}{\bibfnamefont{O.}~\bibnamefont{Monteiro}},
  \bibinfo{author}{\bibfnamefont{M.~R.} \bibnamefont{Pillai}},
  \bibinfo{author}{\bibfnamefont{M.~J.} \bibnamefont{Aziz}}, \bibnamefont{and}
  \bibinfo{author}{\bibfnamefont{M.~C.} \bibnamefont{Ridgway}},
  \bibinfo{journal}{Appl. Phys. Lett.} \textbf{\bibinfo{volume}{82}},
  \bibinfo{pages}{1251} (\bibinfo{year}{2003}).

\bibitem[{\citenamefont{Wang et~al.}(2007)\citenamefont{Wang, Deng, Lu, Sun,
  and Zhao}}]{wang:202503}
\bibinfo{author}{\bibfnamefont{W.~Z.} \bibnamefont{Wang}},
  \bibinfo{author}{\bibfnamefont{J.~J.} \bibnamefont{Deng}},
  \bibinfo{author}{\bibfnamefont{J.}~\bibnamefont{Lu}},
  \bibinfo{author}{\bibfnamefont{B.~Q.} \bibnamefont{Sun}}, \bibnamefont{and}
  \bibinfo{author}{\bibfnamefont{J.~H.} \bibnamefont{Zhao}},
  \bibinfo{journal}{Appl. Phys. Lett.} \textbf{\bibinfo{volume}{91}},
  \bibinfo{pages}{202503} (\bibinfo{year}{2007}).

\bibitem[{\citenamefont{Jaeger et~al.}(2006)\citenamefont{Jaeger, Bihler,
  Vallaitis, Goennenwein, Opel, Gross, and Brandt}}]{jaeger:045330}
\bibinfo{author}{\bibfnamefont{C.}~\bibnamefont{Jaeger}},
  \bibinfo{author}{\bibfnamefont{C.}~\bibnamefont{Bihler}},
  \bibinfo{author}{\bibfnamefont{T.}~\bibnamefont{Vallaitis}},
  \bibinfo{author}{\bibfnamefont{S.~T.~B.} \bibnamefont{Goennenwein}},
  \bibinfo{author}{\bibfnamefont{M.}~\bibnamefont{Opel}},
  \bibinfo{author}{\bibfnamefont{R.}~\bibnamefont{Gross}}, \bibnamefont{and}
  \bibinfo{author}{\bibfnamefont{M.~S.} \bibnamefont{Brandt}},
  \bibinfo{journal}{Phys. Rev. B} \textbf{\bibinfo{volume}{74}},
  \bibinfo{pages}{045330} (\bibinfo{year}{2006}).

\bibitem[{\citenamefont{Zhou et~al.}(2009{\natexlab{a}})\citenamefont{Zhou,
  Shalimov, Potzger, Helm, Fassbender, and Schmidt}}]{zhou_Mn5Ge3}
\bibinfo{author}{\bibfnamefont{S.}~\bibnamefont{Zhou}},
  \bibinfo{author}{\bibfnamefont{A.}~\bibnamefont{Shalimov}},
  \bibinfo{author}{\bibfnamefont{K.}~\bibnamefont{Potzger}},
  \bibinfo{author}{\bibfnamefont{M.}~\bibnamefont{Helm}},
  \bibinfo{author}{\bibfnamefont{J.}~\bibnamefont{Fassbender}},
  \bibnamefont{and} \bibinfo{author}{\bibfnamefont{H.}~\bibnamefont{Schmidt}},
  \bibinfo{journal}{Appl. Phys. Lett.} \bibinfo{pages}{submitted}
  (\bibinfo{year}{2009}{\natexlab{a}}).

\bibitem[{\citenamefont{Bak-Misiuk et~al.}(2009)\citenamefont{Bak-Misiuk,
  Misiuk, Romanowski, Barcz, Jakiela, Dynowska, Domagala, and
  Caliebe}}]{BakMisiuk200999}
\bibinfo{author}{\bibfnamefont{J.}~\bibnamefont{Bak-Misiuk}},
  \bibinfo{author}{\bibfnamefont{A.}~\bibnamefont{Misiuk}},
  \bibinfo{author}{\bibfnamefont{P.}~\bibnamefont{Romanowski}},
  \bibinfo{author}{\bibfnamefont{A.}~\bibnamefont{Barcz}},
  \bibinfo{author}{\bibfnamefont{R.}~\bibnamefont{Jakiela}},
  \bibinfo{author}{\bibfnamefont{E.}~\bibnamefont{Dynowska}},
  \bibinfo{author}{\bibfnamefont{J.}~\bibnamefont{Domagala}}, \bibnamefont{and}
  \bibinfo{author}{\bibfnamefont{W.}~\bibnamefont{Caliebe}},
  \bibinfo{journal}{Mater. Sci. and Eng. B} \textbf{\bibinfo{volume}{159-160}},
  \bibinfo{pages}{99} (\bibinfo{year}{2009}).

\bibitem[{\citenamefont{Shalimov et~al.}(2009)\citenamefont{Shalimov, Potzger,
  Geiger, Lichte, Talut, Misiuk, Reuther, Stromberg, Zhou, Baehtz
  et~al.}}]{shalimov:064906}
\bibinfo{author}{\bibfnamefont{A.}~\bibnamefont{Shalimov}},
  \bibinfo{author}{\bibfnamefont{K.}~\bibnamefont{Potzger}},
  \bibinfo{author}{\bibfnamefont{D.}~\bibnamefont{Geiger}},
  \bibinfo{author}{\bibfnamefont{H.}~\bibnamefont{Lichte}},
  \bibinfo{author}{\bibfnamefont{G.}~\bibnamefont{Talut}},
  \bibinfo{author}{\bibfnamefont{A.}~\bibnamefont{Misiuk}},
  \bibinfo{author}{\bibfnamefont{H.}~\bibnamefont{Reuther}},
  \bibinfo{author}{\bibfnamefont{F.}~\bibnamefont{Stromberg}},
  \bibinfo{author}{\bibfnamefont{S.}~\bibnamefont{Zhou}},
  \bibinfo{author}{\bibfnamefont{C.}~\bibnamefont{Baehtz}},
  \bibinfo{author}{\bibfnamefont{J.}~\bibnamefont{Fassbender}},
  \bibinfo{journal}{J. Appl. Phys.}
  \textbf{\bibinfo{volume}{105}}, \bibinfo{pages}{064906}
  (\bibinfo{year}{2009}).

\bibitem[{\citenamefont{Yamamoto et~al.}(1935)\citenamefont{Yamamoto, Itaya,
  Suga, Takenobu, Iwasa, Hagiwara, Kindo, and Hori}}]{Preisach}
  \bibinfo{author}{\bibfnamefont{F.}~\bibnamefont{Preisach}},
  \bibinfo{journal}{Z.
  Phys.} \textbf{\bibinfo{volume}{94}}, \bibinfo{pages}{277}
  (\bibinfo{year}{1935}).

\bibitem[{\citenamefont{Gottlieb et~al.}(2003)\citenamefont{Gottlieb, Sulpice,
  Lambert-Andron, and Laborde}}]{gottlieb03}
\bibinfo{author}{\bibfnamefont{U.}~\bibnamefont{Gottlieb}},
  \bibinfo{author}{\bibfnamefont{A.}~\bibnamefont{Sulpice}},
  \bibinfo{author}{\bibfnamefont{B.}~\bibnamefont{Lambert-Andron}},
  \bibnamefont{and} \bibinfo{author}{\bibfnamefont{O.}~\bibnamefont{Laborde}},
  \bibinfo{journal}{J. Alloys Compd.} \textbf{\bibinfo{volume}{361}},
  \bibinfo{pages}{13} (\bibinfo{year}{2003}).

\bibitem[{\citenamefont{Luo et~al.}(1991)\citenamefont{Luo, Nagel, Rosenbaum,
  and Rosensweig}}]{PhysRevLett.67.2721}
\bibinfo{author}{\bibfnamefont{W.}~\bibnamefont{Luo}},
  \bibinfo{author}{\bibfnamefont{S.~R.} \bibnamefont{Nagel}},
  \bibinfo{author}{\bibfnamefont{T.~F.} \bibnamefont{Rosenbaum}},
  \bibnamefont{and} \bibinfo{author}{\bibfnamefont{R.~E.}
  \bibnamefont{Rosensweig}}, \bibinfo{journal}{Phys. Rev. Lett.}
  \textbf{\bibinfo{volume}{67}}, \bibinfo{pages}{2721} (\bibinfo{year}{1991}).

\bibitem[{\citenamefont{Sappey et~al.}(1997)\citenamefont{Sappey, Vincent,
  Hadacek, Chaput, Boilot, and Zins}}]{PhysRevB.56.14551}
\bibinfo{author}{\bibfnamefont{R.}~\bibnamefont{Sappey}},
  \bibinfo{author}{\bibfnamefont{E.}~\bibnamefont{Vincent}},
  \bibinfo{author}{\bibfnamefont{N.}~\bibnamefont{Hadacek}},
  \bibinfo{author}{\bibfnamefont{F.}~\bibnamefont{Chaput}},
  \bibinfo{author}{\bibfnamefont{J.~P.} \bibnamefont{Boilot}},
  \bibnamefont{and} \bibinfo{author}{\bibfnamefont{D.}~\bibnamefont{Zins}},
  \bibinfo{journal}{Phys. Rev. B} \textbf{\bibinfo{volume}{56}},
  \bibinfo{pages}{14551} (\bibinfo{year}{1997}).

\bibitem[{\citenamefont{Friedman et~al.}(1997)\citenamefont{Friedman,
  Voskoboynik, and Sarachik}}]{PhysRevB.56.10793}
\bibinfo{author}{\bibfnamefont{J.~R.} \bibnamefont{Friedman}},
  \bibinfo{author}{\bibfnamefont{U.}~\bibnamefont{Voskoboynik}},
  \bibnamefont{and} \bibinfo{author}{\bibfnamefont{M.~P.}
  \bibnamefont{Sarachik}}, \bibinfo{journal}{Phys. Rev. B}
  \textbf{\bibinfo{volume}{56}}, \bibinfo{pages}{10793} (\bibinfo{year}{1997}).

\bibitem[{\citenamefont{Zheng et~al.}(2006)\citenamefont{Zheng, Gu, Xu, and
  Zhang}}]{zhengrk}
\bibinfo{author}{\bibfnamefont{R.~K.} \bibnamefont{Zheng}},
  \bibinfo{author}{\bibfnamefont{H.}~\bibnamefont{Gu}},
  \bibinfo{author}{\bibfnamefont{B.}~\bibnamefont{Xu}}, \bibnamefont{and}
  \bibinfo{author}{\bibfnamefont{X.~X.} \bibnamefont{Zhang}},
  \bibinfo{journal}{J. Phys.-Conden. Matter} \textbf{\bibinfo{volume}{18}},
  \bibinfo{pages}{5905} (\bibinfo{year}{2006}).

\bibitem[{\citenamefont{Song et~al.}(2001)\citenamefont{Song, Roshko, and
  Dahlberg}}]{Song_preisach}
\bibinfo{author}{\bibfnamefont{T.}~\bibnamefont{Song}},
  \bibinfo{author}{\bibfnamefont{R.~M.} \bibnamefont{Roshko}},
  \bibnamefont{and} \bibinfo{author}{\bibfnamefont{E.~D.}
  \bibnamefont{Dahlberg}}, \bibinfo{journal}{J. Phys.-Conden. Matter}
  \textbf{\bibinfo{volume}{13}}, \bibinfo{pages}{3443} (\bibinfo{year}{2001}).

\bibitem[{\citenamefont{Song and Roshko}(2000)}]{Song200024}
\bibinfo{author}{\bibfnamefont{T.}~\bibnamefont{Song}} \bibnamefont{and}
  \bibinfo{author}{\bibfnamefont{R.~M.} \bibnamefont{Roshko}},
  \bibinfo{journal}{Physica B} \textbf{\bibinfo{volume}{275}},
  \bibinfo{pages}{24 } (\bibinfo{year}{2000}).

\bibitem[{\citenamefont{Park et~al.}(1995)\citenamefont{Park, Adenwalla,
  Felcher, and Bader}}]{PhysRevB.52.12779}
\bibinfo{author}{\bibfnamefont{Y.}~\bibnamefont{Park}},
  \bibinfo{author}{\bibfnamefont{S.}~\bibnamefont{Adenwalla}},
  \bibinfo{author}{\bibfnamefont{G.~P.} \bibnamefont{Felcher}},
  \bibnamefont{and} \bibinfo{author}{\bibfnamefont{S.~D.} \bibnamefont{Bader}},
  \bibinfo{journal}{Phys. Rev. B} \textbf{\bibinfo{volume}{52}},
  \bibinfo{pages}{12779} (\bibinfo{year}{1995}).

\bibitem[{\citenamefont{Sun et~al.}(2003)\citenamefont{Sun, Salamon, Garnier,
  and Averback}}]{PhysRevLett.91.167206}
\bibinfo{author}{\bibfnamefont{Y.}~\bibnamefont{Sun}},
  \bibinfo{author}{\bibfnamefont{M.~B.} \bibnamefont{Salamon}},
  \bibinfo{author}{\bibfnamefont{K.}~\bibnamefont{Garnier}}, \bibnamefont{and}
  \bibinfo{author}{\bibfnamefont{R.~S.} \bibnamefont{Averback}},
  \bibinfo{journal}{Phys. Rev. Lett.} \textbf{\bibinfo{volume}{91}},
  \bibinfo{pages}{167206} (\bibinfo{year}{2003}).

\bibitem[{\citenamefont{Sasaki et~al.}(2005)\citenamefont{Sasaki, Jonsson,
  Takayama, and Mamiya}}]{sasaki:104405}
\bibinfo{author}{\bibfnamefont{M.}~\bibnamefont{Sasaki}},
  \bibinfo{author}{\bibfnamefont{P.~E.} \bibnamefont{Jonsson}},
  \bibinfo{author}{\bibfnamefont{H.}~\bibnamefont{Takayama}}, \bibnamefont{and}
  \bibinfo{author}{\bibfnamefont{H.}~\bibnamefont{Mamiya}},
  \bibinfo{journal}{Phys. Rev. B} \textbf{\bibinfo{volume}{71}},
  \bibinfo{pages}{104405} (\bibinfo{year}{2005}).

\end{thebibliography}
\end{document}